\documentclass[12pt]{revtex4}
\usepackage{graphics,epsfig,graphicx}
\usepackage{color}
\usepackage{makeidx}
\usepackage{latexsym}
\usepackage[latin1]{inputenc}

\begin{document}

%======================================%
%<<<<<<<<<<<< TITLE PAGE >>>>>>>>>>>>>>%
%======================================%

\title{Evidence of polariton induced transparency in a single organic quantum wire}

\author{ F. Dubin$^{*}$, J. Berr\'{e}har, R. Grousson, M. Schott, V. Voliotis$^{1}$ \\
{\em Institut des Nanosciences de Paris, UMR 7588 of CNRS} \\
{\em Universit\'{e} Pierre et Marie Curie et Universit\'{e} Denis Diderot,\\
Campus Boucicaut, 140 rue de Lourmel}  {\em 75015 Paris}\\ {\em
$^{1}$ also at Universit\'{e} Evry Val d'Essonne, Boulevard F.
Mitterrand, 91025 Evry cedex}}
\begin{abstract}
The resonant interaction between quasi-one dimensional excitons
and photons is investigated. For a single isolated organic quantum
wire, embedded in its single crystal monomer matrix, the strong
exciton-photon coupling regime is reached. This is evidenced by
the suppression of the resonant excitonic absorption arising when
the system eigenstate is a polariton. These observations
demonstrate that the resonant excitonic absorption in a
semiconductor can be understood in terms of a balance between the
exciton coherence time and the Rabi period between exciton-like
and photon-like states of the polariton.
\end{abstract}

\maketitle

%======================================%
%<<<<<<<<<<<<<<< TEXT >>>>>>>>>>>>>>>>>%
%======================================%

In the 50's Hopfield showed that semiclassical theory is
inappropriate to describe photon absorption by a semiconductor
\cite{Hopfield}. Indeed, an exciton has a defined wave-vector and
is then only coupled to a single photon state, the one having the
same momentum. The resulting system eigenstate during
photoexcitation is a mixed exciton-photon state, a polariton,
which does not lead to photon absorption if additional couplings
are not taken into account \cite{Hopfield}. On the other hand, an
exciton in a quantum well or wire is coupled to a photon continuum
in the emission process. Therefore, the emission probability
follows the Fermi Golden rule (FGR), whereas this is not {\it a
priori} so for the absorption. In fact, as recently shown
\cite{Dubin-Combescot}, two regimes can be reached by
photoexcitation depending on the number of incident photons: a
linear regime where photons are absorbed and a non linear one
where absorption does not occur, the semiconductor becoming
transparent. In the former, the exciton coherence time (generally
governed by interaction with phonons) is much shorter than the
exciton-photon interaction time \cite{Dubin-Combescot}, and
photons are continuously absorbed following the FGR. In the
latter, the system eigenstate is a slightly damped polariton, the
exciton-photon interaction time being comparable to, or shorter
than, the exciton coherence time and absorption saturates. These
considerations allow one to understand why under resonant
excitation, Deveaud and co-workers have observed in a multiple
quantum well Bragg structure a dramatic suppression of the
excitonic absorption \cite{Deveaud}, i.e., the so-called polariton
induced transparency (PIT) regime \cite{Bosacchi}.

In this letter, we report high spatial resolution spectroscopy
(micro-photoluminescence, $\mu$-PL) at low temperature on a single
isolated polydiacetylene (PDA) macro-molecule, a poly-3BCMU red
chain \cite{formule3B}. The elementary optical excitation of this
organic quantum wire being a quasi-1D Wannier exciton, poly-3BCMU
red chains provide an excellent system to investigate the
fundamental interaction between a quasi-1D Wannier exciton and a
photon. This is evidenced by studying a single red chain emission
as a function of excitation power in two situations: firstly, the
excitons are created through pumping within a vibronic absorption
line (non resonant excitation). Secondly, the excitons are
directly created at the band edge (resonant excitation), near
\textbf{k}=0, so they are resonantly coupled to pump photons. In
the former case, neither the $\mu$-PL spectrum nor the chain
excitonic absorption are modified as excitation power increases.
On the contrary, under resonant excitation, we observe a
saturation of the chain excitonic absorption at high excitation
power \cite{pstatsol}. We show that the saturation threshold
occurs at the same excitation power all along the chain spatial
extension (up to 20 $\mu$m) although the chain is only illuminated
over $\approx$ 1 $\mu$m. The observed saturation is interpreted in
terms of a crossover between a regime where photons are absorbed
according to the FGR and the one in which PIT takes place.

A poly-3BCMU quantum wire consists of a chain of Carbon atoms
linked by alternating single, double, and triple bonds, and can be
very diluted in its single crystal monomer matrix. These chains
exist in two electronic structures so-called red and blue phases
\cite{Lecuiller98}. The extremely regular confinement potential
within the 3BCMU monomer crystal leads to an almost ideal chain
spatial conformation, i.e., a perfect one dimensional
semiconductor structure. The optical excitation of poly-3BCMU red
chains is a highly bound (0.6 eV) exciton \cite{Horvath}, which
center of mass motion shows the characteristic pure 1D DOS energy
singularity at the band edge \cite{Dubin2002}. The poly-3BCMU red
chain absorption cross section can be estimated from absorption
measurements of an ensemble of chains \cite{Lecuiller98}, and is
of the order of 10$^{-13}$ cm$^2$ per repeat unit, that is several
orders of magnitude larger than in other known organic or
inorganic semiconductors. The $\mu$-PL spectrum of an isolated red
chain is composed by an intense zero phonon line, due to the
direct recombination of the red exciton, and several much weaker
vibronic replicas deriving from the radiative recombination of the
exciton with coincident emission of a chain optical phonon having
the appropriate momentum \cite{Lecuiller98}. The integrated
intensity of any of these emissions reflects the number of
photocreated excitons along the chain whatever their \textbf{k}
state \cite{Dubin2002}, but only the most intense vibronic
emission, the so-called D-line (since it corresponds to generation
of one double-bound stretch chain optical phonon) is hereafter
considered, all vibronic emissions behaving similarly.

Single chain non resonant excitation (NRE) is performed with the
501.7 nm line of an Ar$^{+}$ laser. The excitation energy is then
within the homogeneous width of the D-phonon assisted absorption
line at low temperature \cite{Lecuiller98}. The zero-phonon and
D-emission lines are recorded as a function of the excitation
power. The resonant excitation (RE) of the excitonic transition is
performed by exciting at the energy of the zero-phonon emission
\cite{Lecuiller98}, and only the D-emission line can be recorded.
The spatial profile of the emission of a single chain is also
presented.

\begin{figure}
\begin{center}
\includegraphics[width=8.5 cm,height=6 cm]{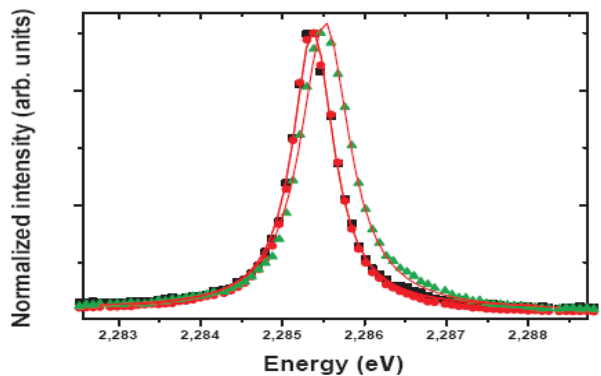}
\caption{Normalized zero-phonon emission under NRE at 10 K for
pump powers of 3.2x10$^{3}$ (filled circles), 2.1x10$^{5}$ (filled
squares), and 1.3x10$^{6}$ W.cm$^{-2}$ (triangles). The lines
correspond to lorentzian fit functions associated with each data.
In the first two cases the FWHM of the lorentzian is 540 $\mu$eV
and its center 2.2854 eV. At 1.3x10$^{6}$ W.cm$^{-2}$, these
parameters are 630 $\mu$eV and 2.2855 eV respectively.}
\label{fig1}
\end{center}
\end{figure}

Figure \ref{fig1} shows the zero-phonon emission line at 10 K as a
function of the NRE power (P$_{NRE}$). Between 3.2x10$^3$ and
6x10$^5$ W.cm$^{-2}$ the emission line-shape remains lorentzian,
centered at 2.2854 eV with a full width at half maximum (FWHM) of
540 $\mu$eV. Then, between 6x10$^5$ and 1.3x10$^6$ W.cm$^{-2}$,
the zero-phonon line broadens and shifts continuously towards
higher energies keeping a lorentzian line-shape. At 1.3x10$^6$
W.cm$^{-2}$ the emission is centered at 2.2855 eV and exhibits a
FWHM of 630 $\mu$eV. According to the temperature dependence of
the $\mu$-PL spectrum of a single red chain \cite{Dubin2002}, the
shift and broadening of the zero-phonon emission with P$_{NRE}$
are shown to both derive from a two Kelvin heating of the chain.
Indeed, for an excitation at 1.3x10$^6$ W.cm$^{-2}$, the
independently measured sample overall temperature has reached
$\approx$12 K. Moreover, the D-emission line-shape remains
unchanged in the whole studied excitation power range (the spectra
can be found in \cite{Dubin2002}).

\begin{figure}
\begin{center}
\includegraphics[width=8.5 cm,height=7 cm]{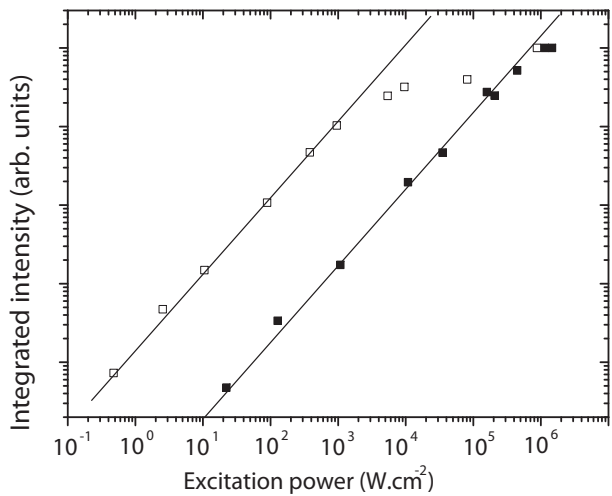}
\caption{Log-log plot of the normalized integrated intensity of
the D-emission line as a function of the excitation power under
NRE (filled squares), and under RE (open squares). The two lines
represent unity slopes, and the data are obtained at 10 and 8 K,
under NRE and RE respectively.} \label{fig2}
\end{center}
\end{figure}

The variation of the D-emission integrated intensity at 10 K
versus P$_{NRE}$ is shown on Figure \ref{fig2} (filled square
symbols). These data show that the number of photocreated excitons
along the chain is proportional to the number of pump photons, and
that exciton-exciton interaction processes are negligible in the
whole studied excitation range. This observation is consistent
with the fact that the excitonic absorption obeys the FGR. Figure
\ref{fig2} also presents the variation of the D-emission
integrated intensity as a function of the RE pump power
(P$_{RE}$). It is proportional to P$_{RE}$ up to $\approx$
4x10$^3$ W.cm$^{-2}$ and then saturates. Indeed, its value for an
excitation at 9x10$^5$ W.cm$^{-2}$ is almost two orders of
magnitude smaller than the one expected if the D-line integrated
intensity remained proportional to P$_{RE}$ in the entire studied
excitation range.

\begin{figure}
\scalebox{0.5}{\includegraphics{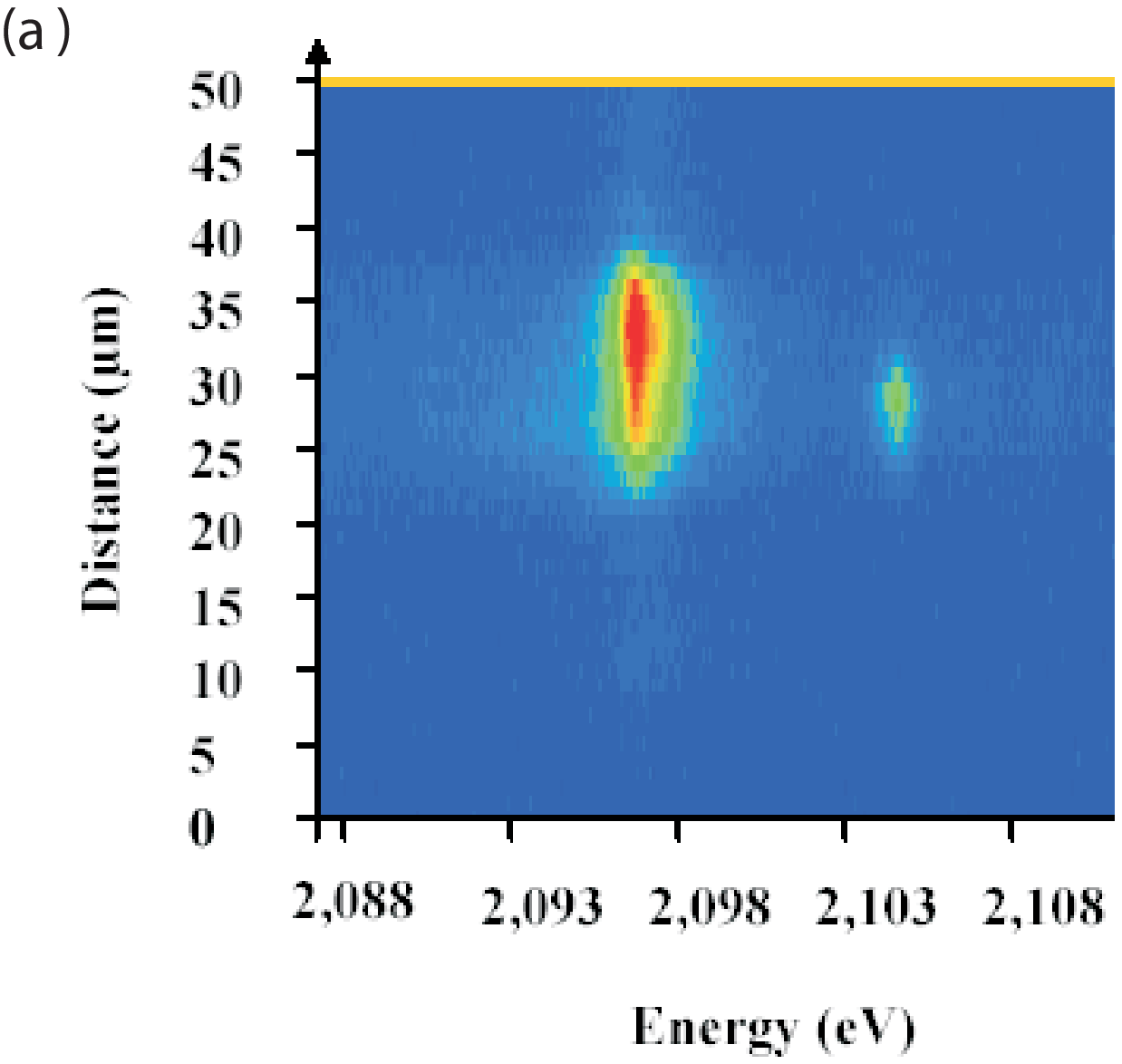}} \hspace*{0.3cm}
\centerline{\scalebox{0.6}{\includegraphics{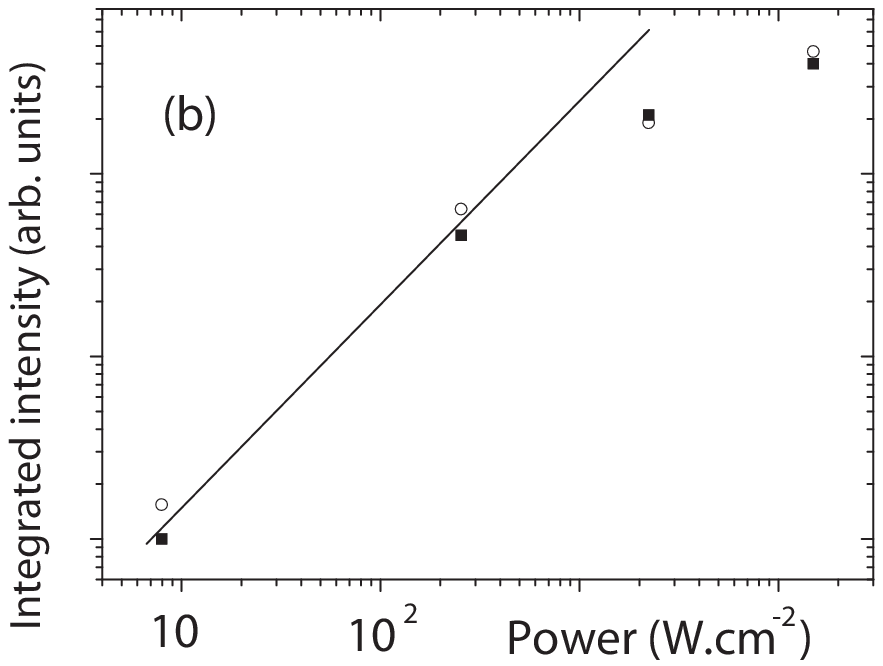}}}
\caption{(a): Image of the D-emission for a RE at 2 10$^3$
W.cm$^{-2}$ and 8 K. The vertical axis is the direction parallel
to the chain, z, graduated in $\mu$m, the horizontal one is the
emission energy in eV. Note that the laser spot position is z=27
$\mu$m with a 2 $\mu$m extension whereas the emission is extended
between z=24 and z=42 $\mu$m. At higher energy a Raman emission of
blue chains is also observed. (b): Variation with P$_{RE}$ of the
D-emission integrated intensity at z=27 $\mu$m (open circles) and
z=37 $\mu$m (squares). The line represents a unity slope.}
\label{fig3}
\end{figure}

Our experimental setup also allows to perform imaging spectroscopy
of a single chain \cite{Guillet2001}. When the red chain axis is
aligned parallel to the spectrometer entrance slits, one obtains
along this direction the spatial extension of the emission, and
the $\mu$-PL spectrum perpendicularly. On Figure \ref{fig3}.a such
an image of a red chain excited resonantly at 8 K is presented. On
this image the vertical axis {\it z}, graduated in $\mu$m,  is the
direction parallel to the chain, the horizontal one is the
emission energy in eV. In this measurement, the laser spot is
centered at z=27 $\mu$m. Therefore, a locally excited single red
chain emits photons from a very extended region, 18 $\mu$m long in
the case of Figure 3.a. The physical origin of this observation
will be addressed elsewhere. The power dependences under resonant
excitation of the D-line emission coming from two separate regions
of this chain, at $z$=27 and $z$=37 $\mu$m respectively, are shown
on Figure 3b. These variations are identical, showing saturation
at the same threshold and the threshold occurs at the same
P$_{RE}$ as on Figure 2, although different chains were studied on
Figure 2 and 3 respectively. Indeed, all individual red chains
show the same saturation behavior.

One possible origin for the observed saturation under RE is a
large exciton population photocreated along the chain. To check
this hypothesis we compare the exciton density created under RE
and NRE. For that purpose the ratio ($\alpha_{NR}/\alpha_R$) for a
single chain must be estimated ($\alpha_{NR}$ and $\alpha_R$ are
the chain absorption coefficients under NRE and RE respectively).
At 13 K, absorption measurements on an ensemble of non interacting
red chains have shown that ($\alpha_{NR}/\alpha_R$)$\simeq$1/10
\cite{Lecuiller98}. The zero-phonon line, corresponding to
resonant absorption, has an inhomogeneously broadened FWHM
$\approx$ 1.5-2 meV; the D absorption line has a FWHM $\approx$
4-5 meV. This width, much larger than the former, is considered to
be a homogeneous width corresponding to the lifetime of the
vibronic state generated by absorption. So, a single isolated
chain D-absorption line should have the same width. It is
independent of temperature up to at least 30 K, the zero-phonon
absorption line area being also independent of temperature, the
ratio ($\alpha_{NR}/\alpha_R$) for a single chain then reads
\begin{equation}\label{xabs}
\left(\frac{\alpha_{NR}}{\alpha_{R}}\right)(T)\simeq\frac{1}{10}\frac{\Gamma_0(T)}{\Gamma_D}
\end{equation}
$\Gamma_0$ and $\Gamma_D$ being the homogeneous linewidths of the
zero-phonon and D absorption line respectively. Around 10 K,
$\Gamma_0$=800 $\mu$eV and $\Gamma_D$=4 meV, the absorption
coefficient under RE is then about 50 times larger than the one
under NRE. From $\alpha_R\simeq$ 10$^{-13}$ cm$^{2}$ we therefore
deduce that for the highest excitation power the dimensionless
exciton density parameter, n$a_X$, n being the exciton density and
$a_X$ the exciton Bohr radius deduced from its polarisability
($a_X$=15$\AA$ \cite{Moller}), is 6x10$^{-2}$ and 2x10$^{-2}$
under NRE and RE respectively \cite{note_nax}. The exciton density
photocreated under NRE being larger than under RE, the saturation
of the chain excitonic absorption cannot be due to exciton-exciton
interactions. Moreover, for such low n$a_X$ values, excitons can
be considered as ideal bosons in the entire excitation range
\cite{Combescot-Tanguy}.

\begin{figure}
\begin{center}
\includegraphics[width=6.5 cm,height=5 cm]{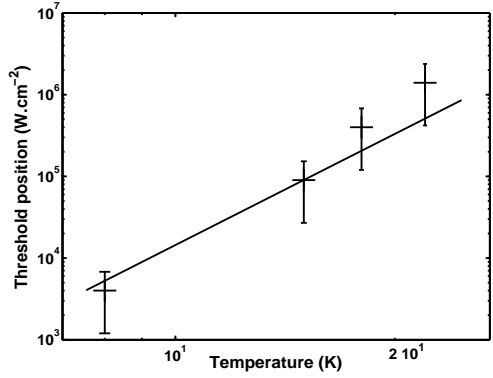}
\caption{Log-log variation of the excitonic absorption saturation
threshold under RE with temperature. The line represents the
$\Gamma_0(T)^3$ predicted behavior.}\label{fig4}
\end{center}
\end{figure}

As mentioned before and precisely shown in \cite{Dubin-Combescot},
the excitonic absorption only follows the FGR when the coherence
time of the exciton state which is coupled to photons, $\tau$, is
short compared to the period of the Rabi oscillations,
$T_N=2\pi/(g\sqrt{N})$, N denoting the number of pump photons and
g the exciton-photon coupling equal to the square root of the
resonant absorption coefficient, $\alpha_R$. The integrated
intensity of the resonant absorption line does not vary with
temperature, the absorption coefficient $\alpha_R$ therefore
scales like (1/$\Gamma_0(T)$) \cite{note_abs} and g like
(1/$\sqrt{\Gamma_0(T)}$). The Rabi period then varies with
temperature like $\sqrt{\Gamma_0(T)}$. Furthermore, for poly-3BCMU
red chains, $\tau$ is governed by the interaction between the red
exciton and the LA phonons of the monomer crystal surrounding the
chain, such that $\tau\propto 1/\Gamma_0(T)$ \cite{Dubin2002}.
Consequently, if the saturation of absorption corresponds to a
transition from a regime in which absorption follows the Fermi
golden rule to another regime where the photon number is large
enough to enter a polariton dominant regime, at the threshold
$\tau\simeq T_N$. The value of $N$ at threshold should therefore
vary with temperature like $\Gamma_0^3(T)$. Indeed, as presented
on Figure \ref{fig4} the measured variation of the saturation
threshold scales like $\Gamma_0(T)^3$ up to at least 20 K
\cite{note_error}. This further supports our interpretation of the
saturated excitonic absorption as an experimental  evidence of a
transition between a regime in which photon absorption obeys the
FGR, to a regime in which PIT takes place.

In fact, when $T_N>\tau$, an exciton initially photocreated at
\textbf{k}=0 is scattered  by LA phonons of the monomer crystal
surrounding the chain to a \textbf{k}$\neq$0 exciton state
\cite{Dubin2002}, before another exciton is created at
\textbf{k}=0. In this regime, very crudely, incident photons are
absorbed one by one and the overall chain excitonic absorption
follows the FGR \cite{Dubin-Combescot}. This first regime
corresponds to the linear variation of the D-emission integrated
intensity with P$_{RE}$. On the contrary, when $T_N<\tau$, the
exciton photocreated at \textbf{k}=0 is strongly coupled to pump
photons such that the system eigenstate is a slightly damped
polariton. This second regime of PIT corresponds to the saturation
of the chain excitonic absorption that we have observed.

To summarize, we have presented the variation of the excitonic
absorption of a single isolated polydiacetylene chain with
excitation power. Under non resonant excitation, neither the
single chain $\mu$-PL spectrum, nor the chain absorption, are
modified as excitation power increases. On the contrary, under
resonant excitation the number of photocreated excitons along the
chain varies proportionally with excitation power and then
saturates. The saturation threshold is shown to be constant all
along the chain spatial extension which is 20 times larger than
the excited part of the chain and to be approximately proportional
to the third power of temperature. The evaluation of the exciton
densities photocreated under resonant and non resonant excitation
shows that excitons behave as ideal bosons and that the saturated
absorption is not due to a too large exciton population along the
chain. The saturation threshold corresponds to a transition
between a regime where photons are absorbed according to the Fermi
Golden rule to the regime of polariton induced transparency. To
conclude let us stress that the polariton induced transparency is
reached thanks to the large absorption cross section of poly-3BCMU
red chains. It requires a very high polariton Rabi frequency which
is proportional to the product of the absorption coefficient and
the number of pump photons. This latter can in our case be kept
sufficiently small to have a low enough excitonic population in
order to avoid effects related to exciton-exciton interactions
which would destroy the bosonic character of excitons and
therefore the polariton induced transparency regime.

\noindent

$^*$: Corresponding author (present adress: Institut für
Experimentalphysik, Universität Innsbruck, Technikerstrasse 25,
A-6020 Innsbruck, Austria)


\begin{thebibliography}{99}
\bibitem{formule3B} The general formula of PDA is
$( =\! CR \! - \! C \! \equiv \! C \! - \! CR' \!=)_{n}$. In
3BCMU, the side groups R and R$'$ are identical, with the
following molecular formula:
-$(CH_{2})_{3}OCONHCH_{2}COOC_{4}H_{9}$
\bibitem{note_abs} The resonant absorption line is a lorentzian of
width $\Gamma_0$, its integral is therefore proportional to
$\alpha_R$$\Gamma_0$.
\end{thebibliography}
\end{document}